**Reassessing Great Britain's Clean Power 2030 Pathway**

**Anthony D Stephens and David R Walwyn**
**Correspondence to tonystephensgigg@gmail.com**

**Abstract**

The National Energy System Operator's (NESO) Clean Power 2030 advice to the government, accepted as United Kingdom (UK) government policy in December 2024, outlines a rapid increase in wind and solar generation capacities with the associated infrastructure by 2030, with the objective of reducing carbon dioxide emissions from the UK's electricity system to 5.2 million tonnes per annum (Mtes). Our previous paper used real-time records from 2017 to analyse NESO's Clean Power 2030 proposal. This paper extends the previous assessment, using real time records of 2017, 2018 and 2019 to test the resilience of the UK energy system under more diverse weather patterns. Two different modelling methods predict that the 2030 carbon dioxide emissions will be five times greater than NESO's Clean Power prediction. Furthermore, the projections show that, due to increased curtailment, only about half of NESO's proposed increases in wind and solar capacities will be economically justified. NESO also advised the government that 35 GW of unabated gas-fired generation would maintain security of supply, whereas our models suggest that at least 42 GW will be required during winter wind lulls. Recent events, such as UK's grid stability problems in 2026, also discussed in the paper, validate our estimated and suggest the risk of blackouts in the UK is increasing.

## Nomenclature

| | |
|---|---|
| GW | Measure of generation in GW |
| $GW_w$ | Wind generation in GW |
| $GW_s$ | Solar generation in GW |
| $GW_{w+s}$ | Wind plus solar generation in GW accommodated by the electricity system |
| Hdrm | That portion of electrical demand available for wind/ solar generation to satisfy |
| IWE | Incremental Wind Efficiency |
| ISE | Incremental solar Efficiency |

## 1. Introduction

The decarbonisation of the United Kingdom's (UK's) electricity system has been a central goal of the country's energy planners over a long period (Or et al., 2027; Dunstan and Drew, 2026). In its most recent planning documents, the National Energy System Operator (NESO) claims that clean power sources will generate 95% of the country's electrical demand in 2030 (NESO, 2024), with the distribution of power and energy per source as shown in Table 1. 35 GW of unabated gas would be maintained to ensure the stability of the electricity system, but would only be used for five percent of annual hours.

**Table 1. NESO's anticipation of total demand and sources of generation (2030)**

| Description | Energy Generation/ Demand (TWh/annum) | Average Annual Power (GW) |
|---|---|---|
| Total UK Demand | 287 | 32.76 |
| Total UK Generation | 358 | 40.87 |
| Offshore Wind | 187 | 21.35 |
| On Shore Wind | 58 | 6.62 |
| Solar Generation | 45 | 5.14 |
| Nuclear | 25 | 2.85 |
| Unabated Gas | 15 | 1.71 |
| Biomass | 15 | 1.71 |
| Exports | 61 | 6.96 |
| Curtailed Generation | 48 | 5.48 |
| Imports | 19 | 2.16 |

Source: Clean Power 2030 Action Plan (NESO, 2024) and the accompanying Technical Annex V_003.

We have previously published a body of independent energy research evaluating the physical feasibility, economic limits, and grid security implications of the UK government's clean electricity decarbonization targets (Stephens and Walwyn, 2024a; Stephens and Walwyn, 2024b; Walwyn and Stephens, 2024; Stephens and Walwyn, 2020). The work specifically challenges the official scenarios produced by the NESO for 2030 and 2035 on the basis that their scenarios take an overly optimistic view of the intermittency of renewable energy and hence underestimate the demand for dispatchable generation. In our work, we have sought to quantify how variable wind and solar generation interact with grid demand and operational headroom, establishing mathematically derived upper deployment limits where curtailment renders additional capacity economically wasteful.

In this article, we extend the analysis by introducing several empirical, methodological, and economic additions that advance our ongoing critique of the UK's Clean Power 2030 strategy and action plan (NESO, 2024; UK Government, 2024). We extend the database of our models to three historical weather years (2017–2019) rather than a single year, the results being independently confirmed by using two entirely different modelling approaches. A new comparison between NESO's 2030 pathway and a lower-renewables case is presented. A recent near failure of the grid in June 2026 is analysed which suggests that the UK cannot rely on imports to maintain grid security.

In Section 2, the updated modelling assumptions and revisions are summarised, and Section 3 presents the results of the new analysis. Section 4 discusses the economic consequences of large wind fleets as proposed by Clean Power 2030, and Section 5 provides commentary on the recent grid failures in Europe and the UK.

## 2. Overview of the Models and Their Assumptions

### 2.1 Assumptions

The values in Table 1 are important inputs to models of the United Kingdom (UK) electricity system in 2030. Our models make three simplifying assumptions:

- Nuclear: Generation will be 2.85 GW throughout the year, leaving an average 29.91 GW available for wind plus solar generation to satisfy. We label this the Headroom, or Hdrm and assume that Hdrm will vary in accordance with variations in electrical demand, the latter being reported by Gridwatch on an real-time basis (Gridwatch, 2025). It is also assumed that unabated gas will create 4.87 Mtes/annum/GW of carbon dioxide emissions, where GW refers to the average annual electrical power output.

- Biomass: The generation level from biomass, as listed in Table 1, is obtained from North American wood pellets at Drax, a former coal fired power station. In 2025 Drax received a subsidy of £1Bn and created 13.3 Mtes of carbon dioxide emissions, equivalent to 7.78 Mtes per GW (Ember, 2025). The burning of wood pellet is only acceptable from an ecological point of view if the resulting carbon dioxide is captured, and that was the original intention at Drax, and the justification for its subsidy. Following the abandonment of the plan to capture carbon dioxide, the government decided that Drax would lose its subsidy in 2027. However, in 2025 the Energy Secretary realised that the 1.51 GW output from Drax would be needed in 2030, and the contract was extended to 2031, with a reduced subsidy of £470M pa (Shanks, 2025). Although the Clean Power 2030 action plan regards Drax generation in 2030 as emission free, independent consultants predict that it will remain the GB's largest emitter of carbon dioxide in 2030 (Ember, 2025). Our models assume that when dispatchable generation is required, whether from gas or biomass generation, there will be carbon dioxide emissions of 4.87 Mtes pa per GW, thereby avoiding the necessity to model biomass generation separately.

- Exports: The Fraunhofer website includes GB in its records of European cross border energy trading and assessed that GB as a net importer of 3.46 GW in 2025 (Fraunhofer, 2026). However, we do not include imports and exports in our real time models, since it is our view, as will be discussed later, that GB's imports and exports are likely to be in balance in 2030.

### 2.2 Formulation of the Models

Historic real time records of the UK electricity system, taken at five minutes intervals, are freely available on the internet (Gridwatch, 2025). These values may be used to assess wind and solar variabilities for the range of weather events likely to be expected during a year and, with appropriate scaling, used to predict the behaviour of future UK electricity systems, assuming the same range of meteorological conditions.

Solar generation is relatively predictable, but this is not the case for wind generation. The wind generation records for 2017, 2018 and 2019 were downloaded a week at a time from the Gridwatch website, the weekly averages for the three years calculated and then normalised to set the annual average generation for each year to 1 so as to allow a direct comparison of wind patterns. The results are shown in Figure 1 and provide a useful means of identifying historic weeks with high and low winds, which are of particular interest when we come to investigation of the likely behaviour of the electricity system should the same weather conditions be experienced in future years.

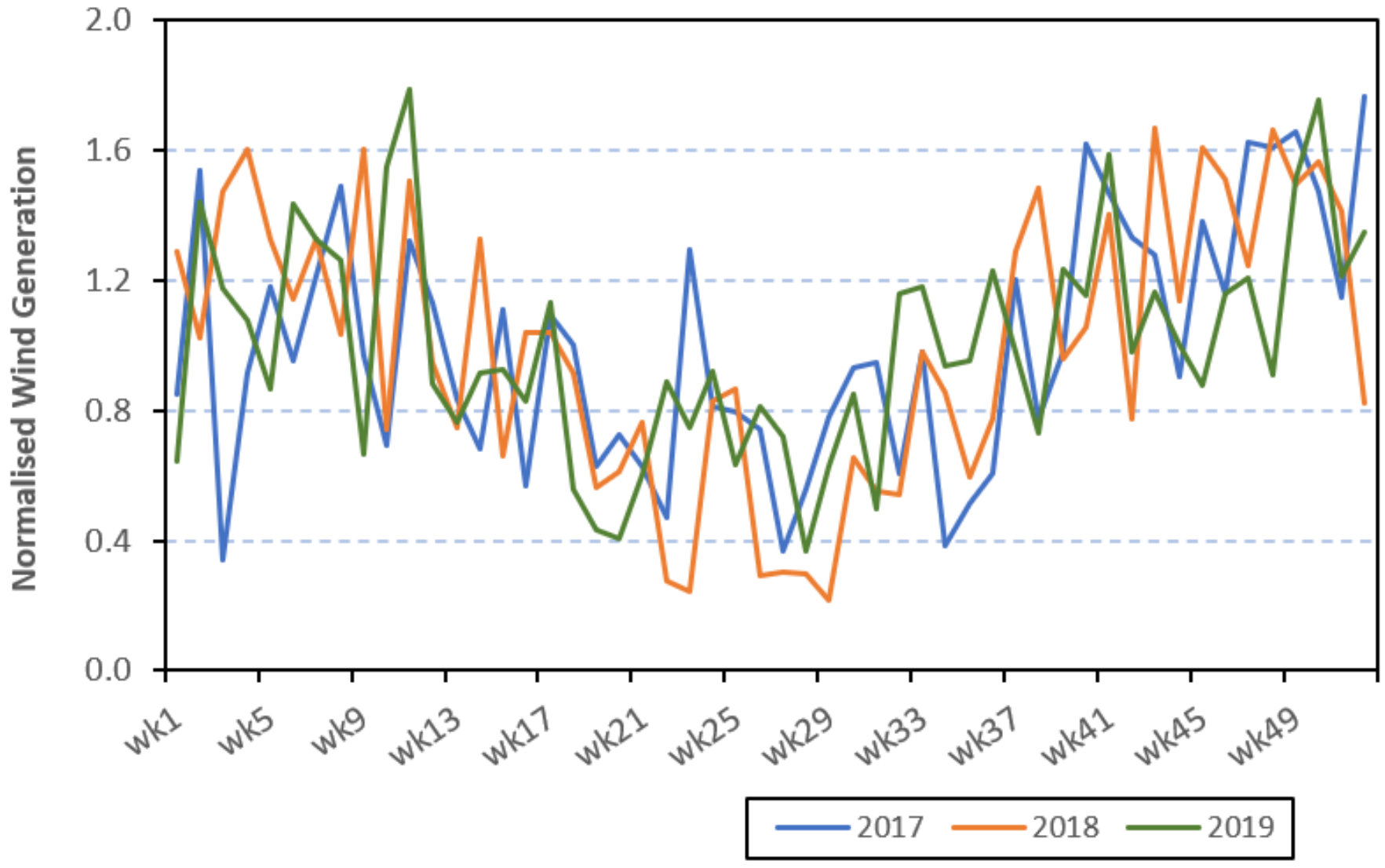

**Figure 1. Normalised weekly wind generations (2017, 2018 and 2019)**

As already noted, it is anticipated that nuclear generation will average 2.85 GW in 2030, leaving electrical demand less 2.85 GW as Hdrm, the latter available to be satisfied by wind and solar generations. The records of electrical demand on the Gridwatch website enable Hdrm to be calculated for each historic time interval, which are then adjusted in accordance with different electrical demands in future years. Suitable scaling of historic wind and solar generations is then applied to calculate their relative contributions to Hdrm in future years for each time interval.

### 2.2.1 Compound Models

The Compound Models comprise 52 weekly models, each of the weekly models being linked to a coordinating model which provides their input variables and collects weekly averages to calculate annual averages. The input variables are Hdrm, annual average wind generation, $GW_w$, and annual average solar generation, $GW_s$. Table 3 is a summary of the Compound Model predictions for 2030 based on the historic records of 2017, 2018 and 2019 respectively.

**Table 2. Modelled parameters for the 2030 electricity system using Compound Models based on the data for 2017, 2018 and 2019**

| Year of Data | 2017 | 2018 | 2019 |
|---|---|---|---|
| Electrical Demand | 32.76 | 32.76 | 32.76 |
| Hdrm | 29.91 | 29.91 | 29.91 |
| $GW_w$ | 27.96 | 27.96 | 27.96 |
| $GW_s$ | 5.14 | 5.14 | 5.14 |
| $GW_{w+s}$ | 33.07 | 33.08 | 33.10 |
| $GW_{w+s}$/Hdrm | 1.10 | 1.11 | 1.11 |
| Accommodated | 24.63 | 24.58 | 24.51 |
| Curtailed | 8.44 | 8.72 | 8.59 |
| Unabated gas | 5.28 | 5.54 | 5.13 |
| Carbon dioxide emissions | 25.72 | 26.99 | 25.0 |
| Incremental Wind Efficiency (IWE) | 0.25 | 0.25 | 0.23 |
| Incremental Solar Efficiency (ISE) | 0.29 | 0.32 | 0.23 |

An extremely useful feature of the Compound Models is that they may be used to calculate the proportion of small increments of GWw and GWs which end up as useful generation accommodated by the electricity system, rather than being curtailed. We call these respectively the Incremental Wind Efficiency (IWE) and Incremental Solar Efficiency (ISE). As may be seen in Table 2, NESO's proposed electricity system configuration for 2030, Hdrm of 29.91 GW, $GW_w$ of 27.96 GW and $GW_s$ of 5.14 GW, results in the incremental efficiencies IWE and ISE falling to 0.25 and 0.29 respectively, based on the 2017 data. This result means that only 25% of the potential generation from the last incremental investments in $GW_w$ and $GW_s$ would be put to beneficial use, 75% being curtailed, and suggests that, for the Hdrm of 29.91 GW anticipated in 2030, it would be more economic to deploy significantly less wind and solar generation than proposed by NESO in Table 1.

The lower efficiencies are even more severe when allowing for the already low proportion of solar or wind output as a proportion of total capacity. For instance, the load factors ($GW_w/GW_{capacity}$) are typically 0.26 for onshore and 0.4 for offshore generators, as shown in Figure 2. The net efficiency is the product of the incremental efficiency and the load factor, which would fall to 6% should IWE reduce to 0.25.

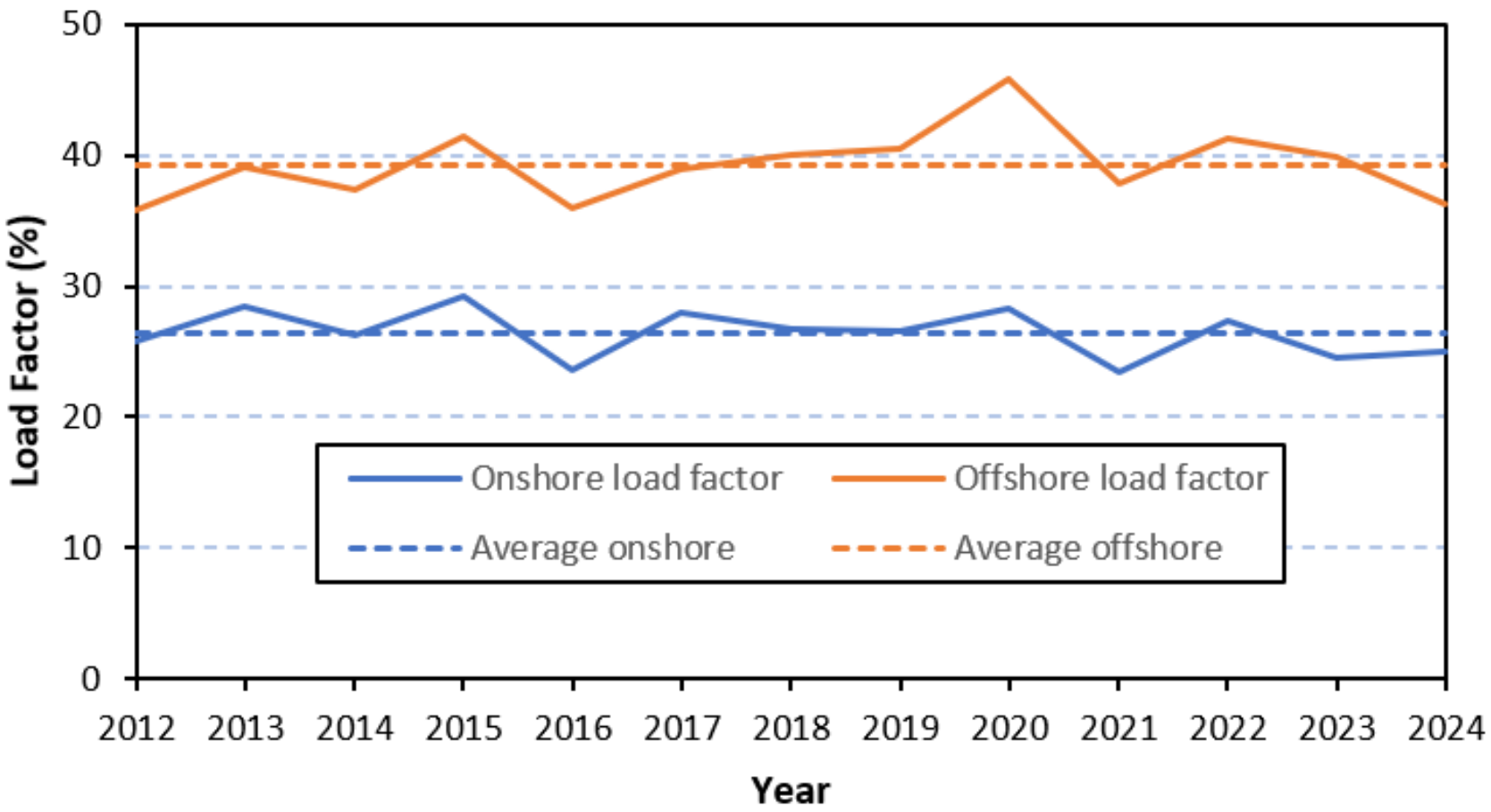

**Figure 2. Typical load factors (efficiencies) for wind generators**

### 2.2.2 Histogram Models

It is highly desirable when modelling complex systems to have alternative computational methods to corroborate model predictions. An alternative method of modelling the performance of future UK electricity systems is to model histograms of annual available wind plus solar generation.

The three blue histograms in Figure 4 are of annual wind plus solar generations in 2030 based on wind and solar generation records of respectively 2017, 2018 and 2019. They were created by first appropriately scaling real time wind and solar generations in 2017, 2018 and 2019, a week at a time, to predict weekly wind and solar generations in 2030. The weekly wind and solar generation predictions are then combined to create weekly wind plus solar generation predictions. The histogram analysis function of Excel was then used to calculate wind plus solar generation histograms for each week, and the weekly histograms were combined to create the blue annual wind plus solar generation histograms of Figure 3. It was then a simple matter to generate the red histograms of wind plus solar generation accommodated by the electricity system.

All generation in generation bands less than Hdrm was accommodated by the electricity system, while only Hdrm was accommodated in generation bands greater than Hdrm. This resulted in the red histograms of generation accommodated by the electricity system in Figure 3. The blue and red histograms then completely specify the electricity system.

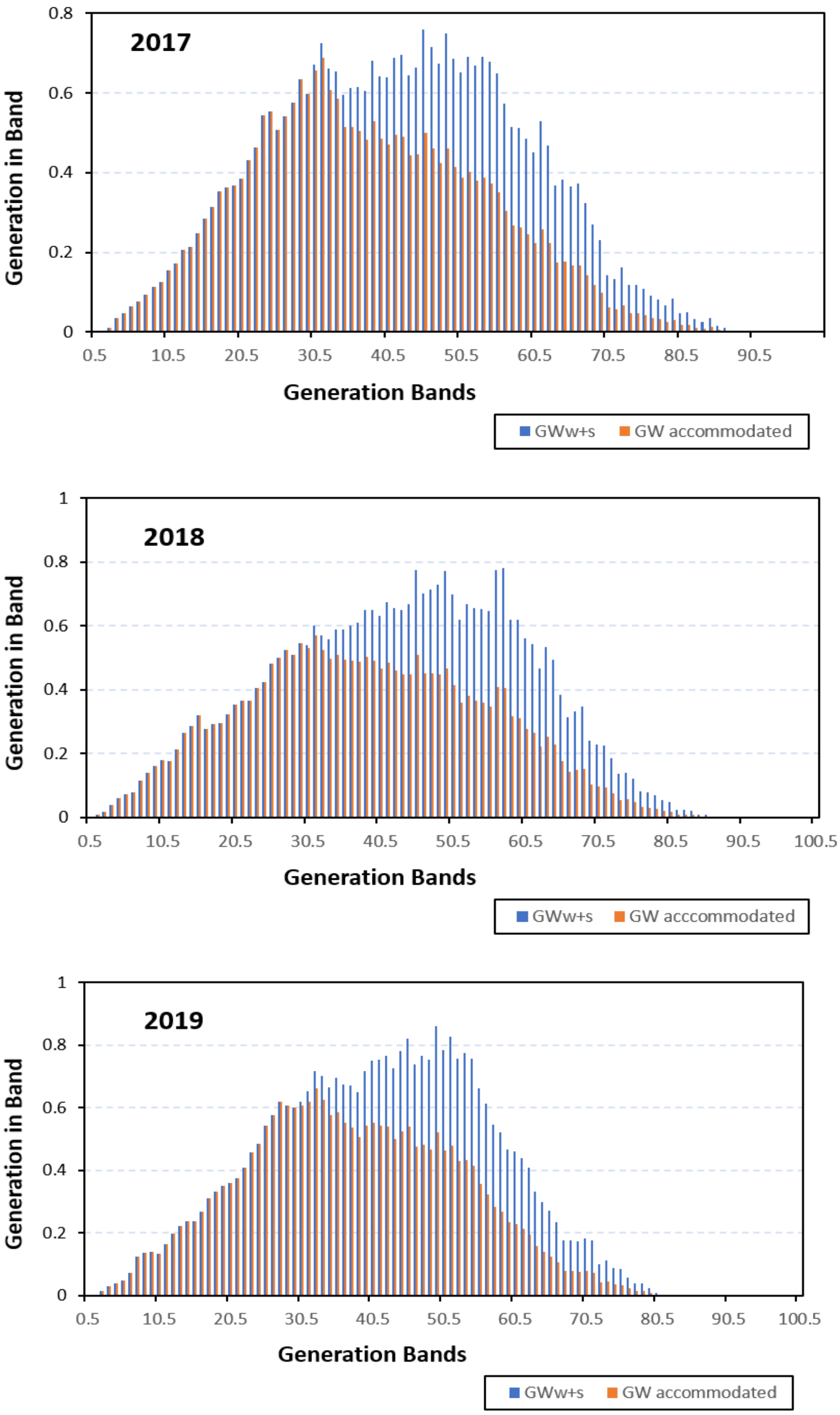

**Figure 3. Histograms of wind plus solar generation in 2030 based on 2017, 2018 and 2019**[1]

[1] The blue histograms are of wind plus solar generation available from the wind and solar fleets and the red histograms are of the generation accommodated by the electricity system

## 3. Results

### 3.1 Model Predictions for 2030 (High Wind Scenario)

Figure 1 shows that week 40 in 2017 (October 2 to 8) was a week of exceptionally high winds, and Figure 4 is the model prediction should the same weather conditions be experienced in 2030. Solar generation is shown stacked on top of wind generation, their combined generations occasionally exceeding Hdrm by up to 50 GW. However, on the last day of the week, wind generation, which had been up to 60 GW for much of the week falls to only 2 GW. Although wind and solar generation are predicted to create 8,271 GWh of energy during the week in 2030, considerably more than the 4,883 GWh needed to satisfy the electricity system, there is a deficit of 364 GWh on the last day of the week.

It is sometimes suggested that stored electrical energy might be used to mitigate wind lulls. However, the plan is for UK to have only 50 GWh of stored electrical energy in 2030 while Europe only plans to have 470 GWh of stored electrical energy. Stored electrical energy would be exhausted in around two hours and its purpose is not to mitigate wind lulls, but rather to work in conjunction with pumped storage to stabilise grid frequency. Stored electrical and pumped stored energies may be seen working together on most days on the internet (Energy Dashboard, 2026). Dispatchable generation would be needed to mitigate the energy deficit of 364 GWh on the last day of the week and would create 0.2 Mtes of carbon dioxide emissions.

This result highlights the need to use dynamic models to calculate carbon dioxide emissions at every time interval to calculate annual carbon dioxide emissions. We shall show later that it would theoretically be possible to export around 4.5 GW of excess generation using UK's 12.5 GW of interconnectors with European countries, if UK's exports were more competitively priced than those in Europe. However, the Fraunhofer website shows that Germany was experiencing similar high winds to those in UK in week 40 of 2017 and was already exporting 10 GW for most of the week (Fraunhofer, 2026). Since Germany plans to have over three times more wind capacity and five times more solar capacity than UK in 2030, it is unlikely that UK would be able to export profitably during a similar week in 2030.

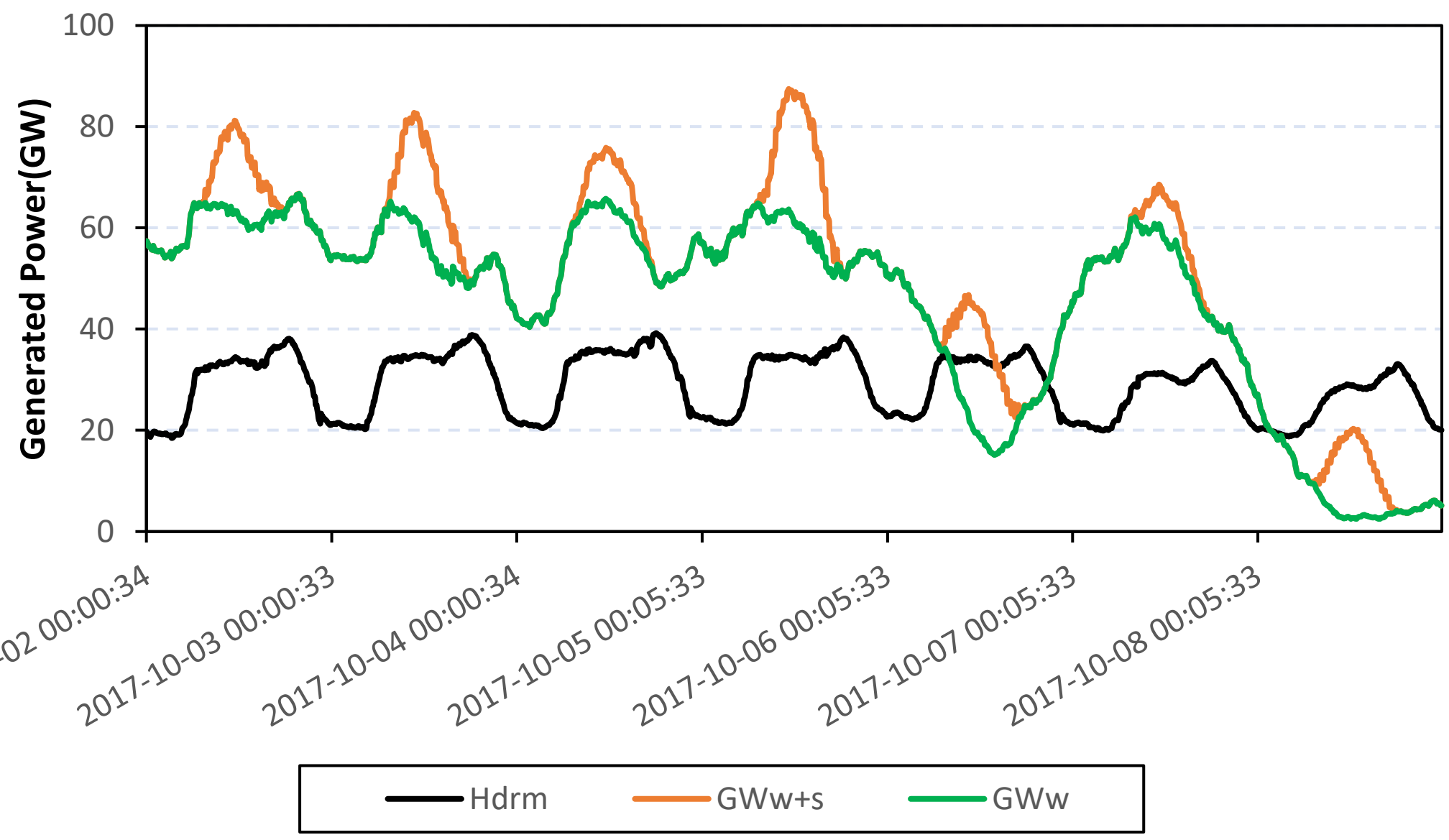

**Figure 4. Model prediction of the UK electricity system in 2030, based on records of week 40 in 2017**

### 3.2 Model Predictions for 2030 (Wind Lull Scenario)

The low average weekly wind generation figures during July of 2018 in Figure 1 reflect the fact that there was an exceptional heat wave during the summer of 2018, not dissimilar to that in 2026 which we shall discuss later. Figure 5 shows the model prediction for 2030, based on records of week 28, July 9-15, in 2018.

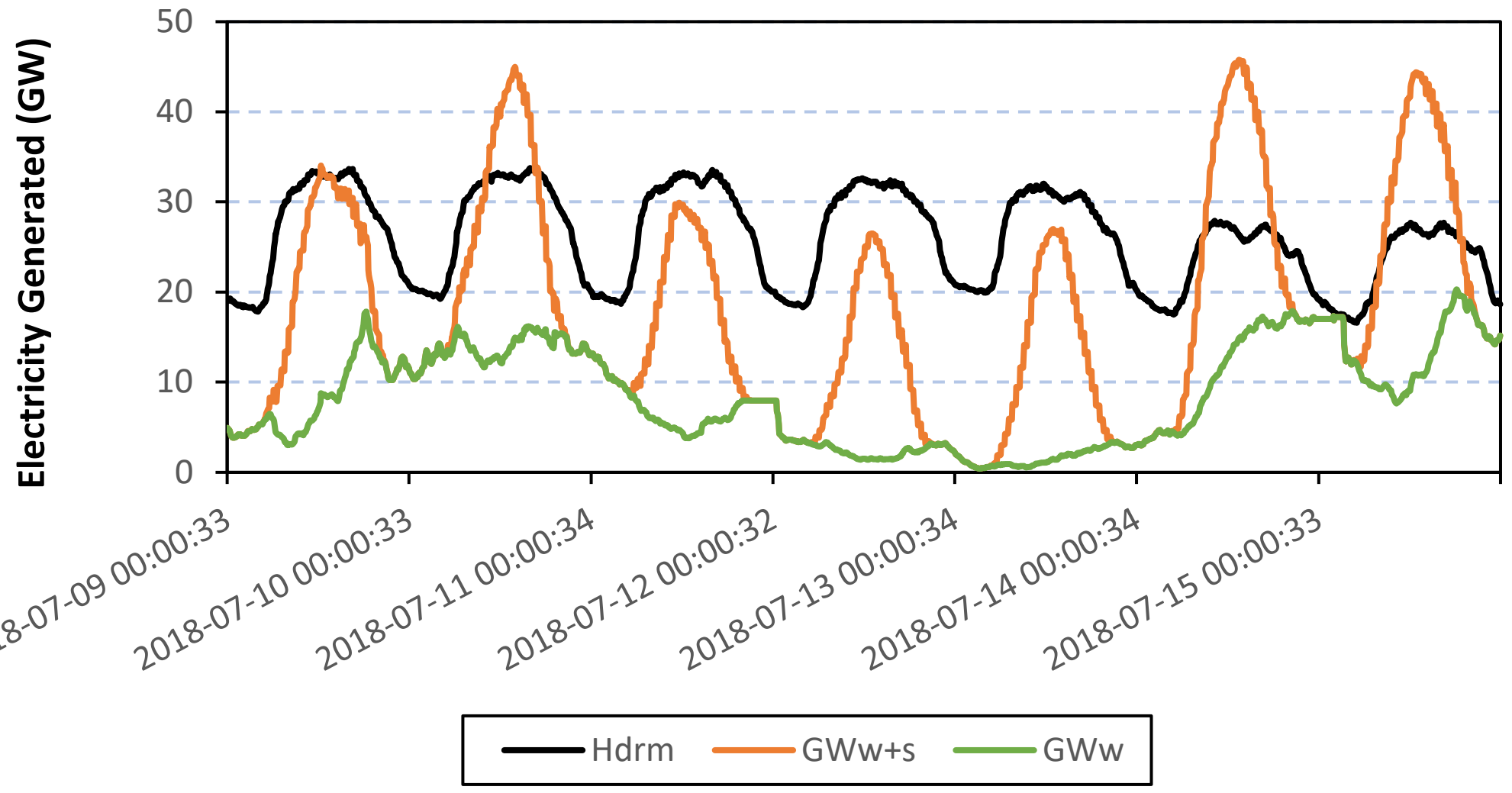

**Figure 5. Model predictions for the UK electricity system in 2030, based on records of week 28, 2018**

The model predictions for the 9 weeks which experienced will lulls of four or more days in 2017, 2018 and 2019 are summarised in Table 3. Although the energy deficits and peak dispatchable generation requirements are more severe during winter wind lulls, Table 3 shows that summer wind lulls are more frequent when there is a heat wave, as in 2018 and 2026. Following the heat wave of 2018 Professor Stephen Belcher, the Met Office Chief Scientist, suggested that although such summer heat waves had previously only occurred every 10 years, *"we might reach a point in future when we can expect a hot summer like that in 2018 to occur every year"* (Met Office, 2018). Later we shall consider a problem of maintaining grid frequency at the height of the heat wave of 2026.

**Table 3. Summary of predictions for 2030 for wind lulls of four or more days in 2017, 2018 and 2019**

| Season | Year | Week No | Dates | Duration of lull (days) | Energy deficit in week (GWh) | Energy deficit (GWh/day) | Peak dispatch (GW) | Carbon dioxide emissions (Mtes) |
|---|---|---|---|---|---|---|---|---|
| Winter | 2017 | 3 | Jan 16-22 | 7 | 4,141 | 592 | 40 | 2.31 |
| Winter | 2019 | 1 | Dec 31-Jan6 | 5 | 2,833 | 566 | 41.6 | 1.27 |
| Summer | 2017 | 27 | July 3-9 | 4 | 1,291 | 322 | 23.5 | 0.72 |
| Summer | 2018 | 22 | May 28-Jn 4 | 5 | 1,897 | 379 | 27.8 | 1.06 |
| Summer | 2018 | 23 | June 4-10 | 5 | 1,957 | 391 | 22.5 | 1.09 |
| Summer | 2018 | 27 | July 2-8 | 5 | 1,394 | 279 | 25.5 | 0.70 |
| Summer | 2018 | 28 | July 9-15 | 5 | 1,618 | 323 | 25.3 | 0.90 |
| Summer | 2018 | 29 | July 16-22 | 6 | 2,090 | 343 | 27.4 | 1.17 |
| Summer | 2019 | 28 | July 8-14 | 4 | 1,453 | 363 | 24.1 | 0.55 |

### 3.3 Exports and Imports

The Compound Models provide a means of calculating how much of UK's excess generation could theoretically be exported using the 12.5 GW of interconnectors with European countries. The 2.16 GW of imports in 2030 is likely to be mainly French nuclear generation transmitted by interconnectors running at a high load factor. This would leave 10.34 GW of interconnector capacity to export UK's excess generation, which otherwise would have to be curtailed. Providing the interconnectors are devoted solely to exporting, the 10.34 GW of interconnectors will increase Hdrm by 10.34 GW. Running the Compound Models with Hdrm of both 29.91 GW and 40.25 GW decreases curtailed generation by the values shown in Table 4, which are the predicted maximum exports of the available 10.34 GW of interconnectors.

**Table 4. Compound Models prediction of potential export of UK excess generation enabled by 10.34 GW of interconnectors**

| Data Years | 2017 | 2018 | 2019 |
|---|---|---|---|
| Export enabled (GW) | 4.56 | 4.58 | 4.39 |

As may be seen in Table 1, NESO anticipates being able to export profitably an average of 6.96 GW in 2030 but Table 4 suggests that the interconnectors would only be able to export 4.4-4.6 GW of UK's excess generation providing the interconnectors are devoted exclusively to exporting UK excess generation to Europe. Although UK will be directly connected to Germany in 2028 when the 1.4 GW NeuConnect interconnector between Kent and Wilhemshaven is commissioned (NeuConnect, 2026; Kastner, 2022), it is already indirectly connected through other countries such as Denmark, Holland, Belgium and Norway with which both countries are interconnected. Germany has been exporting large amounts of excess solar generation to its neighbours for many years, increasingly having to pay its neighbours to take this excess generation. Germany also plans to have significantly more wind and solar generation capacity than UK in 2030, so it seems unlikely that UK will be able to export a significant amount of its excess generation profitably.

This conclusion is consistent with that of independent analysts who concluded that in 2030 79% of the energy flowing down the 1.4 GW NeuConnect interconnector will be from Germany to the UK, only 21% being from the UK to Germany (FTi -CL Energy, 2021). The dynamic model predictions suggest there will be many occasions in 2030 when solar generation far exceeds the needs of the UK electricity system. It highly unlikely however that the excess will provide a profitable export opportunity. Germany, which has been exporting excess solar generation for many years, will have over four times UK's solar capacity and has increasingly had to pay other countries to take its excess. In 2025, Germany legislated to restrict solar access to its distribution network when prices fall below zero (Willis, 2025) and the UK is likely to have to take similar action before 2030.

Our modelling assumption is that in 2030 UK's imports and exports are both likely to be around 2 GW. It is not actually possible to model the European energy balances and prices, since both are continually changing. However, if UK imports and exports are balanced in 2030 at around 2 GW, they may, for modelling purposes, be ignored. Wind plus solar generation which the Compound Models calculate to exceed UK's needs are assumed to be curtailed.

### 3.4 Histogram model predictions for 2030

The results of the histogram model are summarised in Table 5. It is assumed that an annual GW of unabated gas generation results in 4.87 Mtes pa of carbon dioxide.

**Table 5. Histogram model predictions for the electricity system in 2030 based on 2017, 2018 and 2019 records, compared to Clean Power 2030**

| Data Year | 2017 | 2018 | 2019 | Clean Power 2030 |
|---|---|---|---|---|
| Electrical demand | 32.76 | 32.76 | 32.76 | 32.76 |
| Hdrm | 29.91 | 29.91 | 29.91 | 29.91 |
| GWw | 27.6 | 27.6 | 27.6 | 27.6 |
| $GW_s$ | 5.14 | 5.14 | 5.14 | 5.14 |
| GWw+s | 33.03 | 33.13 | 33.16 | 33.11 |
| Accommodated | 24.27 | 23.76 | 24.49 | 27.63 |
| Curtailed | 8.86 | 9.37 | 8.41 | 5.48 |
| Unabated gas | 5.64 | 6.15 | 5.42 | 1.71 |
| Carbon dioxide emissions | 27.47 | 29.95 | 26.39 | 5.2 |

Since the Compound Models predictions of carbon dioxide emissions in 2030 in Table 5 took account of the variations in Hdrm throughout the year, between about 20 GW in summer and 40 GW in winter, we cannot expect the Histogram Model which assumes a constant Hdrm of 29.91 GW to give identical predictions. However, the fact that the Histogram and Compound Models predictions agree with each other to within 5-10% gives good confidence in both modelling approaches.

## 4. Economic Consequences of Large Wind Fleets

In Section 3 we saw that NESO's proposed deployment of wind and solar generations in 2030 will lead to over 25% of available wind plus solar generation being curtailed, and that only 25% of the last increments of wind and solar generations would be accommodated by the electricity system and used to decarbonise the electricity system. The upper economic limit of deployment of wind and solar generations are likely to be defined by the lowest economically acceptable IWE and ISE values and Figure 6 summarises the three Compound Model predictions of IWE and carbon dioxide emissions for a range of $GW_w$ values from 0 to 65 GW, with $GW_s$ held constant at 5.14 GW. It may be seen that the predictions of carbon dioxide emissions based on 2017 and 2019 records are almost identical, while the predictions based on 2018 records only diverge a little for large $GW_w$ values which are of no practical interest.

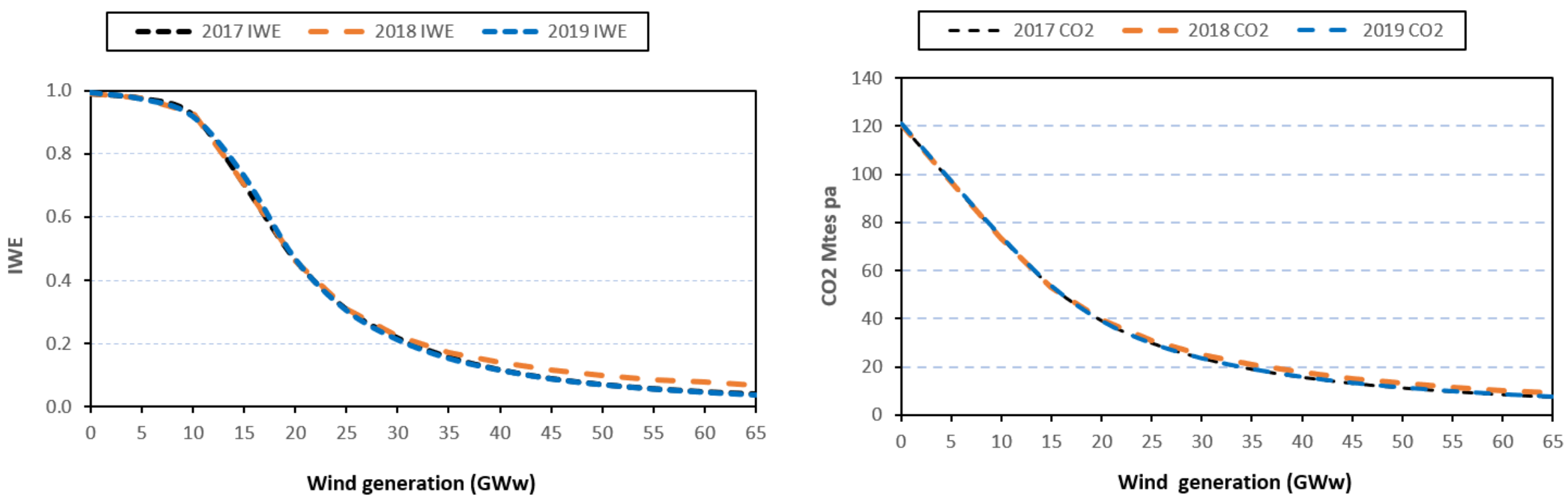

**Figure 6. Compound Model predictions of carbon dioxide emissions (left) and IWE values (right) based on records**

Figure 6 shows that a progressive decrease in the efficiency with which wind generation decarbonises the electricity system at around 20 GW (IWE = 0.5). The Compound Model provides a means of calculating IWE and ISE values for any Hdrm, $GW_w$ and $GW_s$ values, and an earlier investment planning analysis showed how this process might be reversed, i.e. calculating the $GW_w$ and $GW_s$ values which satisfy Hdrm and equal values of IWE and ISE (Stephens and Walwyn, 2024b). It is possible to extend and generalized this analysis in terms of the non-dimensional ratios, $GW_w$/ Hdrm, $GW_s$/Hdrm and $GW_{w+s}$/Hdrm as shown in Table 6.

**Table 6. Relationships between $GW_w$/Hdrm, $GW_s$/Hdrm $GW_{w+s}$/Hdrm and IWE=ISE**

| IWE=ISE | 0.2 | 0.3 | 0.4 | 0.5 | 0.6 | 0.7 |
|---|---|---|---|---|---|---|
| GWw/Hdrm | 1.013 | 0.838 | 0.727 | 0.646 | 0.578 | 0.521 |
| GWs/Hdrm | 0.212 | 0.186 | 0.166 | 0.15 | 0.138 | 0.127 |
| GWw+s/Hdrm | 1.226 | 1.024 | 0.893 | 0.796 | 0.716 | 0.648 |

Column 3 in 7 shows the $GW_w$ and $GW_s$ values calculated using Table 6 for Hdrm=29.91 and IWE=ISE=0.5. Confirmation of the validity of the calculated $GW_w$ and $GW_s$ values is the closeness of the recalculated IWE and ISE values to 0.5 using the Compound Model, shown at the bottom of column 3 in Table 7. In moving from the 2024 configuration in column 2 to column 3, 2.5 GW of the additional 12.83 GW deployed would be curtailed. IWE=ISE=0.5 implies that half of a further increment of either $GW_w$ or GWs would be curtailed, and that the marginal cost would have risen by a factor of two. In moving from column 3 to column 4, 5.9 GW however of the increase of 9.28 GW would be curtailed.

IWE and ISE values falling to around 0.25 implies that the marginal cost of a further incremental deployed would have risen by a factor of 4. Since detailed costs of the 2030 project are not in the public domain, it is not possible to calculate an economic optimum ratio of GWw and GWs values in 2030. However, much of the cost of the £200Bn Clean Power project is associated with wind and solar infrastructure, so it seems likely that increasing wind plus solar generation from 10.98 GW in 2024 to 23.81 GW in 2030 would be considerably more economic than increasing wind plus solar generations to 33.09 GW. This conclusion is reinforced by EdF's publicly expressed concerns about infrastructure costs turning out to be two or three times greater than budgeted for in the Clean Power 2030 plan, and its request that Ofgem and NESO make costs transparent and carry out a strategic review of the Clean Power 2030 project (Buckland and Ball, 2026).

**Table 7. Comparison of scenarios**

| Scenario | 2024 | 2030 IWE/ISE=0.5 | NESO 2030 |
|---|---|---|---|
| Hdrm | 25.74 | 29.91 | 29.91 |
| $GW_w$ | 9.47 | 19.32 | 27.95 |
| $GW_s$ | 1.51 | 4.49 | 5.14 |
| $GW_{w+s}$ | 10.98 | 23.81 | 33.09 |
| $GW_{w+s}$/Hdrm | 0.48 | 0.796 | 1.10 |
| Accommodated $GW_{w+s}$ | 10.98 | 21.28 | 24.64 |
| Curtailed | 0.02 | 2.52 | 8.45 |
| Dispatch generation required | 14.79 | 8.82 | 5.28 |
| Carbon dioxide (Mtes pa) | 72.05 | 42.04 | 25.70 |
| IWE | 0.97 | 0.51 | 0.25 |
| ISE | 1 | 0.50 | 0.29 |

## 5. Recent Stability Events in Energy Systems

The UK government's target is for renewable sources to provide 95% of UK's electricity in 2030 but Bialek has pointed out that there is insufficient operational experience of systems powered by such high percentages of variable renewable energy to be confident they can be operated securely (Bialek, 2026). A useful starting point for identifying some of the more important security issues associated with renewable energy may be found in Bialek's analyses of a UK grid failure in 2019 (Bialek, 2020), a failure of the Iberian grid in 2025 (Bialek, 2026), and problems in ensuring UK grid security in 2026.

### 5.1 United Kingdom Grid Failure (2019)

The initial cause of the UK power outage on 9 August 2019, which affected over a million residential customers, and rail travel into the following day, was a lightning strike on a Transmission Network in East Anglia. Lightning strikes on Transmission Networks are not infrequent, and reserve generating capacity should have been sufficient to mitigate the loss of any single source of generation on the network. However, what caused the system to fail was the subsequent loss of three different sources of generation, 737 MW loss of generation from the Hornsea offshore wind farm, 244 MW loss of output at the Little Barton combined cycle gas power station, and 500 MW loss of "embedded" generation attached to the Distribution Network. Bialek (2020) suggested that these failures were a consequence of control system inadequacies:

- a discrepancy between Hornsea's onshore controller and individual controllers on its offshore wind turbines
- discrepancies between the three independent speed controllers on the steam turbine at Little Barton power station which caused the steam turbine to trip
- the tripping of the Distribution Network, with loss of its 500 MW of "embedded" generation, when the Transmission Network frequency fell to 49 Hz, even though the trip setting for the Distribution Network should have been set to 47 Hz.

A conclusion which may be drawn from the UK grid failure in 2019 is that control of electricity systems powered by renewables is becoming increasingly challenging because of the increasing number of systems which must be controlled. This is particularly the case for the Distribution Networks to which a very large number of relatively small solar generating units are being attached, but none of which are visible to the Systems Operator.

### 5.2 Iberian Grid Failure (2025)

A major failure of the Spanish electricity system on 28th April 2025 caused 60 million people on the Iberian Peninsula to be without power for 24 hours. Bialek provides useful insightful into both the causes of the grid failure, and the lessons which might be learnt from what was the first major black-out caused by over-voltage (Bialek, 2026). The blackout was preceded by several voltage oscillations on the Transmission Network. It was not these oscillations which caused of the blackout, rather an increasing voltage on the Transmission Network which was not satisfactorily controlled. Bialek identifies several reasons for the increasing voltage escalating into a full black-out, including:

- a lower than usual number of conventional plants available on the Transmission Network, some of which did not control voltage as planned
- few conventional plants being located to the south of Spain where most of the renewable generation is located. Bialek also cited examples of unexplained voltage oscillations of 8Hz in Northern Scotland in 2021 and 2023, where there is a large amount of wind generation but little conventional generation (Bialek, 2024)
- overvoltage protection systems being set at too low a level, and therefore tripping when it was unnecessary to do so

- renewable generation plants operating at constant power factor, their outputs not being controlled to match the phase of the grid. He contrasted Spain's approach to inverter design unfavourably with the approach in UK, Ireland and Portugal, where inverters providing power from renewable sources matched the phase of the grid
- Spain had the most ambitious renewable energy program in Europe, but had not, in Bialek's opinion, matched its enthusiasm for increasing renewable generation with sufficient attention to the consequences for the operation of the resulting system.

In its final report on the Spanish power failure, ENTSO-E, the body which coordinates the operation of Europe's interconnected power grid, pointed out that voltage control was becoming more challenging. It demanded that network codes should be revised and that every kind of generation should be subject to dynamic voltage control (In Power, 2026)

### 5.3 United Kingdom Grid Instability (2026)

On 23rd June 2026, Gridwatch records reveal that grid frequency fell below the critical lower level of 49.8 Hz in the early evening, and imports, which had been significant earlier in the day, fell to zero. A "Significant Event Report", written by NESO immediately after the incident, recorded that "the system constraints were breached, and system security was compromised…. Frequency repeatedly dropped below operational limits between about 7pm and 8pm, thanks to high demand during the evening peak, low wind generation, changes in flows through international cables, and failure of expected imports to arrive" (Ellery and Samuel, 2026).

To understand why imports were unavailable to UK on this evening is necessary to study the records available on the Fraunhofer website of cross border imports/ exports, for both European countries and UK. Although the cross border records for many European countries are available, only those which were major net exporters (positive in Table 8) and major net importers (negative in Table 8) had a significant impact on UK's ability to import. Column 2 shows average net cross border trading for the year 2025, and column 3 the average net cross border trading for 23rd June 2026. It may be seen that in 2025 the main net exporters were France (10.56 GW), with its substantial nuclear fleet, and Sweden (4.0 GW) and Norway (2.71 GW) with their hydroelectric power. Germany was no longer a net importer following of closure of its nuclear reactors and reduction of coal fired generating capacity, and the UK was a net importer because of the closure of its older nuclear reactors and all its coal fired power stations.

**Table 8. European and UK most significant electricity importing/ exporting countries in 2025**

| Country | Average net trading in 2025 (GW) | Net trading on 23rd June 2026 (GW) |
|---|---|---|
| France | 10.56 | 6.32 |
| Sweden | 4.0 | 1.43 |
| Norway | 2.71 | 2.30 |
| Spain | 1.06 | 3.35 |
| Total of main net exporters | 18.32 | 13.39 |
| Germany | -2.51 | -6.24 |
| UK | -3.46 | -1.00 |
| Italy | -5.58 | -5.10 |
| Total of main net importers | -11.55 | -12.35 |

Table 8 reveals that on 23rd June 2026 the main electricity exporting countries were exporting around 5 GW less than their average in 2025, mainly as a consequence of a number of French nuclear reactors having to shut because of high river temperatures (Crellin and Lhomendet, 2026) and reduced hydroelectric power in Scandinavia because of low rainfall the previous winter. The Gridwatch and Energy Dashboard websites show that at mid-day on 23rd June UK was importing 7.74 GW from Norway, Denmark, Holland, Belgium and France, but by 7pm these imports had disappeared. The Fraunhofer website reveals that the exports from the countries which had exported to UK at mid-day had switched to Germany in the evening. At midday, Germany's 60.47 GW of solar generation had resulted in an excess of 19.95 GW, but this excess disappeared as solar generation fell in the afternoon. By 8pm (7pm BST) Germany was importing 12.7 GW, 6.47 GW from the countries which had been exporting to the UK at mid-day. The shortage of surplus electricity generation in Europe in the evening caused the price to rise to 773 Euro/MWh.

What is perhaps surprising is that NESO expected to able to import on the evening of 23rd June 2026. The reduced export abilities of France and Scandinavia were well publicized in the Press, and the Fraunhofer website's day-ahead predictions of electrical demand and generation had accurately predicted the magnitude of its deficit in the evening. It may be concluded from this analysis that the UK should not assume it will always be able to import from its neighbours, but should always have sufficient dispatchable generation available in reserve to mitigate adverse meteorological conditions such as wind lulls. Our models suggest peak demands for dispatchable generation of up to 42 GW in winter and up to 28 GW in summer (see Table 3).

NESO's assumption is that 35 GW of unabated gas powered generation would always be available to mitigate wind lulls, so it is of some concern that it was only able to deploy 17.6 GW during the wind lull on 23rd June 2026. A possible explanation for the lack of gas generating capacity is that most of the fleet was built in the early to mid- 1990's and must be coming to the end of its useful life (Porter). Also, since the government's strategy has been to reduce gas fired generation there has been little incentive for private generating companies to further invest in gas generation. The companies will have been aware that the Clean Power 2030 strategy assumed that the existing 35 GW fleet would have a load factor of only 5 % in 2030, and NESO's energy flow diagram in its Future Energy Scenario for 2022 shows stored hydrogen rather than natural gas being used to mitigate wind lulls. Table 8 shows that the UK imported an average of 3.46 GW in 2025 and it seems likely that the problems of maintaining grid stability on 23rd June 2026 was a consequence of the decommissioning of 3.95 GW of generation capacity in 2022 (2GW of coal fired generation at Ratcliffe- on- Sour, 1GW of nuclear generation at Hinkley Point B and 1GW of nuclear capacity at Hunterston B).

An immediate concern following the electricity system's difficulty in coping with a summer wind lull of 23rd June is how it will be able to cope with a winter wind lull, when electrical demand will be considerably greater than in summer. Of longer term concern is that it has been estimated that a further 12 GW of firm capacity will be taken out of service by 2030 (Porter, 2015). It is possible that such considerations were the reason that for Ofgem authorising NESO on 10th August to implement rolling 3 hour power cuts on a regional basis at 8 hours' notice, should it be necessary to do so to avoid more widespread blackouts (Barltrop, 2026; Farhat, 2026; Oliver, 2026).

Similar incidents of instability occurred on 6th October and 16th October. UK and German records show both countries experiencing wind lulls, NESO only able to deploy 18.45 GW of unabated gas and issuing an emergency notice of a possible shortfall of 1.9 GW. Indeed, it has been reported that there had been concern about power stability on three different days during the previous week and Andrew Larkin, a former Ofgem advisor, suggested that the "risks of blackout appeared to be increasing" (Samuel et al., 2026).

## 6. Conclusion

The three sets of Compound Model predictions based on real time historic records of respectively 2017, 2018 and 2019 suggest that carbon dioxide emissions in 2030 will be 25-27 Mtes. Confidence in these predictions is increased by the Histogram Model predictions, based on the same records, which suggest that carbon dioxide emissions in 2030 will be in the range 26-30 Mtes. These predictions are about five times higher than NESO's advice to the UK government in 2024 that carbon dioxide emissions in 2030 will be 5.2 Mtes.

We are unable to compare modelling approaches, since NESO's models are proprietary. However, the Clean Power Action Plan 2030 advice to government includes the information that the advice was based on a Dynamic Dispatch Model designed for the Dept of Energy and Climate Change in 2011. This model was designed to assess investment options for dispatchable sources of generation, which in 2011 were predominantly coal and gas fired. We would question whether a model designed to analyse investment options for dispatchable generations will have a structure which enables it to analyse the performance of electricity systems mainly powered by variable renewable energy. We would further suggest that the dynamic performance of systems powered predominantly by variable renewable energy can only be assessed using models based on real time historic records.

The analysis in Section 4 suggests that the efficiency with which wind and solar generations can decarbonise the UK electricity system will decrease progressively as they increase relative to the headroom available for them to satisfy. Our models predict that the Clean Power 2030 Action Plan implies increasing wind and solar generations to the point at which the marginal cost will be four times greater than the marginal cost without curtailment. We suggest that it is likely to be more economic to increase wind plus solar generation, which was 10.98 GW in 2024, to 23.81 GW in 2030, rather than to 33.09 GW as proposed by NESO, particularly since EdF suggests that the infrastructure investments needed to transmit the additional renewable generations are already significantly delayed, and costs are turning out to be significantly higher than anticipated in 2024 (Buckland and Ball, 2026).

A major concern for electricity systems powered mainly by renewables is how they will cope with lengthy wind lulls. The consequences for wind lulls in 2030, based on the wind lulls of four or more days experienced in 2017, 2018 and 2019 are discussed in Section 3.2. The wind lull experienced in week 26 of 2026 was similar to that in Figure 3, based on the wind lull of week 28 2018, both weeks experiencing near record heat waves. It is surprising that the Energy Dashboard/ Historical records show that only 17.6 GW of gas generation being available for deployment on the evening of 23rd June 2026, and only 18.45 GW on October 6th since NESO advised 35 GW would be maintained to mitigate wind lulls. Since the Meteorological Office predicts that heat waves are becoming more frequent, a lesson from 23rd June 2026 is that UK must not rely on imports during wind lulls but must always have sufficient dispatchable generation available to satisfy UK's electrical demand. In Section 5 a possible explanation is suggested for the limited availability of gas generation in 2026. Also discussed are the consequences for coping with future wind lulls of the anticipated loss of 12 GW of firm generating capacity by 2030.

The failures of the UK grid in 2019 and the Spanish grid in 2025 highlight the consequences of inadequate controllers and incorrect trip settings. Bialek (2026) expresses concern that there is insufficient experience of systems powered mainly by renewables to be certain that they will be able to operate securely, and ENTSO suggests every kind of generation should be subject to dynamic voltage control. An increasing amount of UK generation, particularly solar generation, is being attached to the Distribution Networks and is therefore not visible to the System Operator. To avoid future grid failures, there must be a strong case for all sources of UK generation, whether on the Transmission or Distribution Networks, being monitored and controlled, as suggested by ENTSO for the European networks.